\documentclass[aps,prl,twocolumn]{revtex4}
\usepackage{color}
\usepackage{graphicx}
\usepackage{amsmath}
\usepackage{amssymb}
\usepackage{epstopdf}
\begin{document}

\title{Vertically-Illuminated, Resonant-Cavity-Enhanced, Graphene-Silicon Schottky Photodetectors}

\author{M. Casalino$^1$, U. Sassi$^2$, I. Goykhman$^2$, A. Eiden $^2$, E. Lidorikis$^3$, S. Milana$^2$, D. De Fazio$^2$, F. Tomarchio$^2$, M. Iodice$^1$, G. Coppola$^1$, A. C. Ferrari$^2$}

\affiliation{$^1$National Research Council, Institute for Microelectronics and Microsystems, 80131 Naples,Italy\\
$^2$Cambridge Graphene Centre, University of Cambridge, 9 JJ Thomson Avenue, Cambridge, CB3 0FA, UK\\
$^3$Department of Materials Science and Engineering, University of Ioannina, Ioannina 45110, Greece\\}
\begin{abstract}
We report vertically-illuminated, resonant cavity enhanced, graphene-Si Schottky photodetectors (PDs) operating at 1550nm. These exploit internal photoemission at the graphene-Si interface. To obtain spectral selectivity and enhance responsivity, the PDs are integrated with an optical cavity, resulting in multiple reflections at resonance, and enhanced absorption in graphene. Our devices have wavelength-dependent photoresponse with external (internal) responsivity$\sim$20mA/W (0.25A/W). The spectral-selectivity may be further tuned by varying the cavity resonant wavelength. Our devices pave the way for developing high responsivity hybrid graphene-Si free-space illuminated PDs for free-space optical communications, coherence optical tomography and light-radars.
\end{abstract}
\maketitle
\section{Introduction}
Near infrared (NIR) photodetection at 1550nm is of paramount importance for a variety of applications, ranging from optical communications\cite{Agra2010,LeitJOFCR2005, KausIEEE2016,Ijaz2012} to remote sensing\cite{FlooPERS2001,ParriAGU2011}. In modern telecom systems, operation at 1550nm benefits from a reduced light absorption in optical fibers\cite{Agra2010}. In free-space optical communications (FSO)\cite{LeitJOFCR2005, KausIEEE2016, Ijaz2012} and light-radars (LIDARs)\cite{FlooPERS2001,ParriAGU2011}, this minimizes the propagation losses in fog\cite{KausIEEE2016} and humid conditions\cite{Ijaz2012}, due to lower optical absorption and scattering compared to wavelengths$<1\mu$m, and improves eye safety because the outer layer of the eye (cornea) absorbs light at 1550nm and does not allow it to focus on the retina\cite{FlooPERS2001,KausIEEE2016}. In optical coherence tomography (OCT), a non-invasive imaging technique for biological tissues\cite{IshiBOE2012}, the advantages of using 1550nm are enhanced penetration depth, due to lower scattering in tissue with respect to shorter wavelengths\cite{IshiBOE2012}, and enhanced imaging contrast at deeper penetration depths, where multi-scattering processes dominate\cite{IshiBOE2012}.

Many photodetectors (PDs) for 1550nm have been proposed\cite{Agra2010, Sze2006, EngN2015, CasalJOA2012}. For telecom and datacom applications, these typically rely on a waveguide configuration\cite{KangNP2009, MasiAPL2002, KoesIEEE2006, MichNP2010, AsseN2010, ZhuAPL2008, GoykNL2011, GoykOE2012, CasaJAP2013, AkbaOE2010}, in which optical confinement and guiding contribute to enhanced light absorption and photodetection. On the other hand, for FSO, OCT and LIDARs, NIR PDs for free-space illumination are required\cite{LeitJOFCR2005, KausIEEE2016, Ijaz2012, FlooPERS2001, ParriAGU2011, IshiBOE2012}. At present, III-V compound (\textit{e.g.} InGaAs, InP)\cite{WaldIEEE1996, KangIEEE2002} and group IV (Ge)\cite{KangNP2009, MasiAPL2002, KoesIEEE2006, MichNP2010} semiconductors are the materials of choice for vertically-illuminated NIR PDs, due to their high ($>$90\%)\cite{Sze2006} NIR absorption. The ever growing demand and performance requirements in modern systems (such as bit-rate, number of pixels, imaging matrix size, operation and processing speed)\cite{Agra2010,Sze2006} make it crucial to integrate PDs with supporting circuitry (drivers, amplifiers, processors) on the same chip. Since modern microelectronics relies on mature complementary metal-oxide-semiconductor (CMOS) technology, the development of NIR PDs on Si is promising for integrated microsystems, combining both optical and electronic functionalities. III-V materials are not compatible with standard CMOS fabrication processes because of cross-contamination and dopant redistribution effects\cite{SzeVLSI2003}, and are typically manufactured in separate facilities\cite{SzeVLSI2003}. They can be later bonded with CMOS chips using advanced packaging and assembling\cite{Lau1999}. However, the overall performance can degrade due to packaging parasitics (\textit{e.g.} parasitic capacitance and inductance) and cross-talks associated with the wire-bond leads\cite{Lau1999}. Epitaxially grown Ge on Si provides a competitive platform to InGaAs and InP based NIR photodetection\cite{KoesIEEE2006, HaraASS2004, WangS2011}. Nevertheless, due to defects\cite{SzeVLSI2003} and dislocations-like recombination centers at the Si-Ge interface\cite{KoesIEEE2006, HaraASS2004, WangS2011}, these PDs typically show larger leakage current\cite{WangS2011} and smaller shunt resistance\cite{WangS2011} compared to III-V devices, resulting in increased noise (thermal and shot)\cite{KoesIEEE2006}. To reduce the defects density, a two-step Ge deposition is commonly used\cite{KoesIEEE2006, HaraASS2004, WangS2011}, however it involves high ($>$650$^\circ$C) temperature processes\cite{WangS2011} that can sacrifice the thermal budget, which limits the amount of thermal energy available to the wafer during the fabrication process.

An alternative and promising approach for monolithic integration of NIR PDs with CMOS electronics is to perform sub-bandgap photodetection in Si exploiting the internal photoemission process (IPE) in a Schottky junction\cite{Sze2006,PetePIEEE1962,BronNN2015,CasalJQE2016}. In this case, photoexcited carriers from the metal electrode can be emitted to Si over the Schottky barrier $\Phi_B$, allowing NIR detection for photon energy h$\nu>\Phi_B$\cite{Sze2006,PetePIEEE1962,BronNN2015}. Schottky PDs have been successfully used in IR focal plane arrays (FPA)\cite{KimaOR1998}, monolithically integrated with CMOS readout electronics and charge coupled devices (CCDs)\cite{Sze2006}. The advantages of the Schottky junction configuration over other PD types (pn and pin junctions, quantum wells) stem from its simple structure\cite{Sze2006}, easy fabrication and integration with CMOS technology\cite{Sze2006}. The main drawback is the limited ($<1\%$)\cite{GoykA2014, CasaS2016} internal quantum efficiency (IQE) of the IPE process, defined as the number of carriers emitted to Si per absorbed photon. This is mainly because the momentum mismatch between the electron states in the metal and Si results in a specular reflection of excited electrons at the Schottky interface\cite{Sze2006,GoykA2014}. The IQE is linked to the PD internal (external) responsivity R$_{int}$ (R$_{ext}$), defined as the ratio between the photocurrent I$_{ph}$ and the absorbed (incident) optical power $P_{abs} (P_{inc})$, \textit{i.e} $R_{int}$= $I_{ph}/P_{abs}$= $IQE\cdot q/h\nu$ and $R_{ext}$=$I_{ph}/P_{inc}$=$A{\cdot}R_{int}$\cite{Sze2006}, where \textit{q} is the electron charge and \textit{A} is the absorptance. As a result, limited IQE leads to limited responsivity, so that the highest R$_{ext}$ reported so far in vertically-illuminated Si Schottky PDs operating at 1550nm is$\sim$5mA/W\cite{DesiO2015}, much lower than the 0.5-0.9A/W for III-V\cite{Sze2006} and Ge\cite{Sze2006} based PDs. Graphene/Si Schottky PDs at 1550nm have been demonstrated both in free-space\cite{AmirIEEE2013} and guided mode configurations\cite{WangNP2013,GoykNL2016}, with $R_{ext}$ up to 10mA/W and 0.37A/W respectively. In these devices, a single layer graphene (SLG) acts as electrode in contact with Si, forming a Schottky junction with rectifying characteristics\cite{ChenNL2011, AnNL2013, BartoPR2016}. In general, graphene is an attractive material for photonics and optoelectronics\cite{FerrN2015,KoppNN2014,BonaNP2010,SunACS2010}. Its integration with Si may allow the development of miniaturized and cost-effective hybrid optical devices and functionalities\cite{SorianCM2017}. In the case of SLG/Si Schottky PDs, SLG integration allows absorption of NIR photons with energy below the Si bandgap in close proximity to the Schottky interface\cite{AmirIEEE2013,AnNL2013,WangNP2013,GoykNL2016}, leading to increased IPE IQE. However, the SLG absorption in NIR is$\sim$2.3${\%}$\cite{NairS2008, DawlAPL2008}, and the vast majority of optical power does not contribute to photodetection. As a result, vertically-illuminated SLG/Si Schottky PDs have a limited R$_{ext}\sim$10mA/W\cite{AmirIEEE2013} at 1550nm, over one order of magnitude lower compared to R$_{ext}\sim$0.37A/W\cite{GoykNL2016} for the waveguide-integrated configuration, in which light is fully absorbed in SLG upon optical guiding.

Here we increase R$_{ext}$ of free-space illuminated graphene/Si Schottky PDs by combining a Schottky junction with an optical Fabry-Perot (F-P) cavity to enhance light interaction and absorption at the SLG/Si interface. We show that the PD spectral response and responsivity peaks coincide with F-P resonances, with R$_{ext}$ increasing with the number of light round trips inside the cavity. Taking advantage of multiple ($\sim$5) light reflections at resonance, we obtain spectrally-selective photoresponse with maximum R$_{ext}$ (R$_{int}$)$\sim$20mA/W (0.25A/W), the highest reported so far for vertically-illuminated Si PDs at 1550nm. Our device paves the way for high responsivity hybrid graphene/Si PDs for NIR.
\begin{figure}
\centerline{\includegraphics[width=85mm]{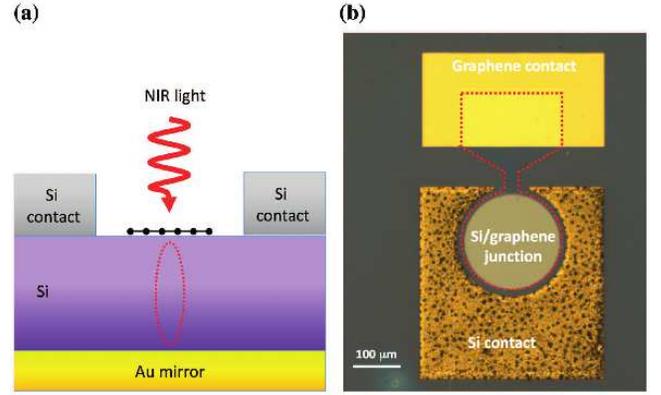}}
\caption{(a) Schematic cross-sectional view of resonant cavity enhanced (RCE) SLG/Si Schottky PD under illumination. (b) Optical image.}
\label{fig:Fig1}
\end{figure}
\begin{figure}
\centerline{\includegraphics[width=85mm]{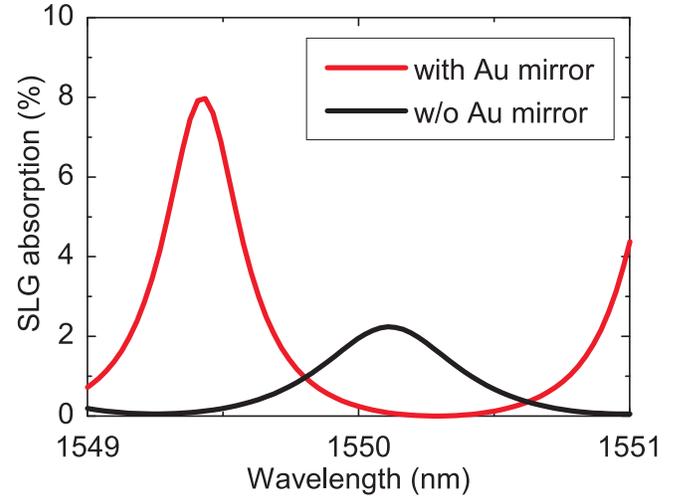}}
\caption{Calculated SLG absorption on top of a cavity with (red) and without (black) Au mirror.}
\label{fig:Fig2}
\end{figure}
\begin{figure}
\centerline{\includegraphics[width=85mm]{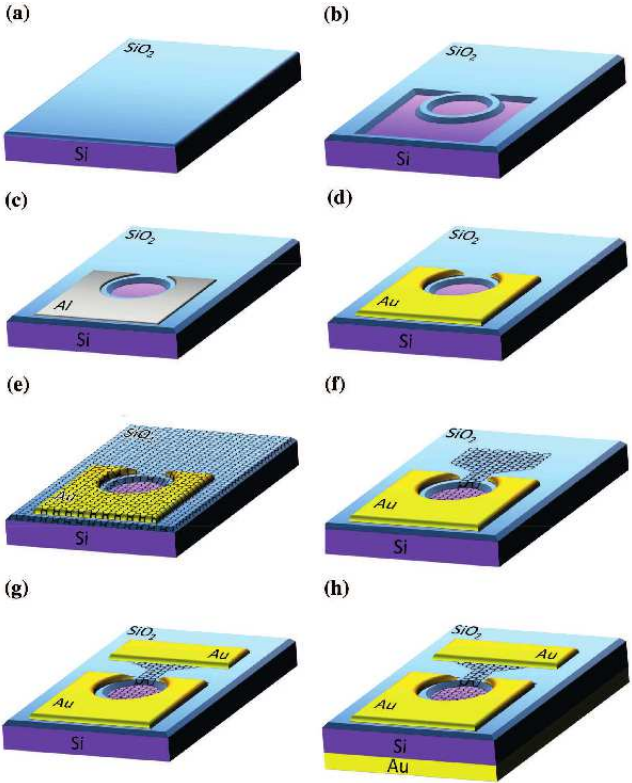}}
\caption {Fabrication process of RCE Si-SLG Schottky PD. (a) SiO$_2$ layer deposition. (b) Schottky and Ohmic contacts area definition. (c) Al ohmic contact formation. (d) Au protection layer deposition. (e) SLG transfer. (f) SLG shaping. (g) Au contact to SLG deposition. (h) Au Back mirror deposition.}
\label{fig:Fig3}
\end{figure}
\begin{figure}
\centerline{\includegraphics[width=85mm]{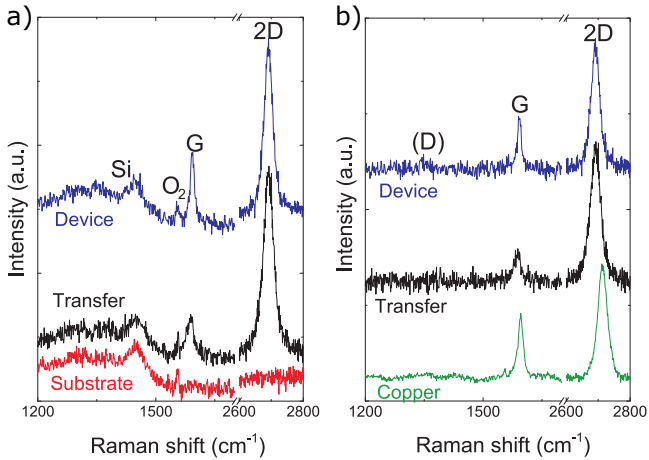}}
\caption{(a) Raman spectra of (red) Si substrate, (black) SLG transferred on Si and (blue) of SLG on Si after device fabrication. (b) Raman spectra after subtraction of the substrate contribution of (green) as-grown SLG on Cu, (black) SLG transferred on Si and (blue) SLG on Si after device fabrication}
\label{fig:Fig4}
\end{figure}

Fig.\ref{fig:Fig1} illustrates our F-P cavity integrated SLG/Si Schottky PD. The resonant structure consists of a $\lambda/2$ Si slab layer confined between SLG/Si top and Au bottom mirrors. When vertically-illuminated at resonance, light circulates inside the cavity leading to increased absorption at the SLG/Si interface, resulting in enhanced IPE from SLG to Si. Fig.\ref{fig:Fig2} shows a simulation of SLG absorption in the integrated F-P cavity PD. For this we use the transfer matrix method\cite{MuriIEEE1997} and a cavity length \textit{L}=200$\mu$m. The Au permittivity $\epsilon_{Au}$=-115+11.3i (at 1550nm) is assumed to follow the Drude model including damping\cite{JCPRB1972}. The permitivity of SLG $\epsilon_{SLG}$=1.8+16.4i (at 1550nm) is calculated from its optical conductivity $\sigma$, \textit{i.e.} $ \epsilon=1+i\sigma/(\omega \cdot \epsilon_0 \cdot \Delta)$, where $\omega$ is the angular frequency, $\epsilon_0$ is the permitivity of vacuum, and $\Delta$=0.35nm is the SLG thickness. $\sigma$=61-4.3i [$\mu$S] includes both contributions from the inter- and intraband transitions\cite{Falk&Pers2007,Falk&Varl2007}, assuming a SLG doping$\sim$0.25eV and scattering time$\sim$50fs. The simulation indicates that the absorption peaks are spaced by 1.65nm (Fig.\ref{fig:Fig2}), matching the free-spectral range (FSR) of the F-P cavity\cite{Agra2010}: FSR=$\lambda_0^2/(2\cdot n_g \cdot L)$, where $\lambda_0$ is the illumination wavelength, $n_g=n - \lambda_0 \cdot dn/d \lambda$ is the group index of Si around 1550nm, estimated to be $n_g\sim$3.61\cite{LiJPC1993}, n=3.45 is the refractive index of Si at 1550nm\cite{LiJPC1993}, and $dn/d \lambda$ is the dispersion. We get peak absorptances$\sim$2.2 and 8.5$\%$ for finesse (\textit{i.e} ratio between FSR and resonance linewidth) of 3 and 5.3 without and with Au mirror, respectively (Fig.\ref{fig:Fig2}). The peak absorptance for the larger finesse (Au mirror) is$\sim$4 times higher that for the lower finesse (Si/air mirror), due to larger number of light round trips and optical energy build-up inside the F-P cavity.

The device fabrication process is presented in Fig.\ref{fig:Fig3}. We use a double-polished, low-doped ($\sim$10$^{15}$cm$^{-3}$), 200$\mu$m thick Si substrate to minimize the scattering and free-carriers losses in the F-P cavity. First, a 100nm thick SiO$_2$ layer is deposited by e-beam evaporation, Fig.\ref{fig:Fig3}a, then patterned by optical lithography using a laser-writer (Microtech), followed by wet etching of SiO$_2$ in a buffer-oxide-etch (BOE) solution, Fig.\ref{fig:Fig3}b. Next, Al ohmic contacts to the p-type Si are realized by an additional lithographic step, followed by lift-off and alloying at 460$^\circ$C for 30min in a forming gas (5.7$\%$ H$_2$ in N$_2$), Fig.\ref{fig:Fig3}c. To protect the Al pads from subsequent treatments involving HF, we cover Al with an Au layer using optical lithography, Cr/Au (3nm/50nm) evaporation and lift-off, Fig.\ref{fig:Fig3}d. The ohmic contact imperfections (black spots) in Fig.\ref{fig:Fig1}b arise from the Al/Si alloy process.

SLG is grown by chemical vapor deposition (CVD) on a 35-$\mu$m-thick Cu foil, following the process described in Ref.\citenum{BaeNN2010}. The quality of the material is monitored by Raman spectroscopy using a Renishaw InVia system equipped with a 100$\times$ objective at 514.5nm and a laser power below 300$\mu$W. Fig.\ref{fig:Fig4}b (green curve) shows the spectrum of SLG on Cu, after the removal of the background Cu photoluminescence (PL)\cite{LagatAPL2013}. The two most intense features are the G and the 2D peak, with no significant D peak. The 2D peak is single-Lorentzian, signature of SLG\cite{FerrPRL2006}. The position of the G peak, Pos(G), is$\sim$1595cm$^{-1}$, with FWHM(G)$\sim$9.5cm$^{-1}$. The 2D peak position, Pos(2D) is$\sim$2712cm$^{-1}$  with FWHM(2D)$\sim$29cm$^{-1}$. The 2D to G peak intensity and area ratios, I(2D)/I(G) and A(2D)/A(G), are$\sim$1.9 and 5.8, respectively, suggesting a p-doping$\sim$400meV\cite{FerrNN2013, DasNN2008, BrunACS2014}. After growth, the SLG is wet-transferred to the target substrate, Fig.\ref{fig:Fig3}e. The film is coated with 500nm polymethyl methacrylate (PMMA), followed by Cu etching in ammonium persulfate (APS). The resulting SLG/PMMA film is rinsed in water to remove APS residuals. To obtain a SLG/Si Schottky interface without a native oxide layer, we transfer SLG in diluted HF in deionized (DI) water (HF/DI water; 1:100)\cite{GoykNL2016}. After removing the APS residuals, the SLG/PMMA layer is placed in a plastic beaker containing 5mL/500 mL HF and DI water. Next, the target substrate is first dipped in a buffered oxide etch (BOE) for 5s to etch the native Si oxide and then used to lift the floating SLG/PMMA layer. As a result, during drying, HF at the SLG/Si interface prevents Si oxidation\citep{GoykNL2016}. After drying, the sample is placed in acetone to dissolve the PMMA, leaving the SLG covering the target substrate, Fig.\ref{fig:Fig3}e. After transfer, we use additional optical lithography steps to shape the SLG by oxygen plasma and then deposit Cr/Au (3nm/50nm) contacts by evaporation, followed by lift-off, Fig.\ref{fig:Fig2}f. Before evaporation, a mild (0.5W, 20s) Ar plasma (Moorfield NanoETCH) is applied on the exposed SLG areas for PMMA residuals cleaning leading to low ($\sim$100$\Omega$) contact resistance at the metal-SLG interface that was separately estimated using a transfer length method . Finally, a bottom mirror is realized by 100nm-thick-Au layer, thermally evaporated on the backside, Fig.\ref{fig:Fig3}h. Fig.\ref{fig:Fig1}b shows an optical image of a representative device, where the SLG boundaries are highlighted by a red dashed line. The SLG quality after transfer and after the complete fabrication process is also monitored by Raman spectroscopy. The Raman spectrum of SLG after transfer is shown in Fig.\ref{fig:Fig3}b, black curve. This is obtained by point-to-point subtraction of the reference Si spectrum (Fig.\ref{fig:Fig4}a, red curve) from the transferred SLG (Fig.\ref{fig:Fig4}a, black curve), when the intensities in both spectra are normalized to the third order Si peak at $\sim$1450cm$^{-1}$ \citep{TempPRB1973}. The 2D peak after transfer is a single-Lorentzian, Pos(2D)$\sim$2691cm$^{-1}$  with FWHM(2D)$\sim$35cm$^{-1}$. Pos(G) is 1587.5cm$^{-1}$ with FWHM(G)$\sim$19.6cm$^{-1}$. The 2D to G peak intensity and area ratios, I(2D)/I(G) and A(2D)/A(G), are 4.32 and 7.16 respectively, suggesting a p-doping$\sim$2.4$\cdot$10$^{12}$ cm$^{-2}$ ($\sim$200meV)\cite{FerrNN2013,DasNN2008, BaskPRB2009}. Fig.\ref{fig:Fig4}a (blue curve) plots the SLG Raman spectrum after device fabrication and point-to-point subtraction of the Si reference using the same procedure. After device fabrication Pos(G)=1592.4cm$^{-1}$, FWHM(G)$\sim$10.7cm$^{-1}$, Pos(2D)$\sim$2691.4cm$^{-1}$ FWHM(2D)$\sim$31.1cm$^{-1}$. I(2D)/I(G) and A(2D)/A(G) are 2.53 and 6.3 respectively, indicating doping$\sim$9$\cdot$10$^{12}$cm$^{-2}$ ($\sim$290meV)\cite{FerrNN2013, DasNN2008,BaskPRB2009}. We also get I(D)/I(G)$\sim$0.14, indicating that the fabrication process does introduce significant defects in the SLG electrode\citep{CancNL2011,BrunACS2014}.

\begin{figure}
\centerline{\includegraphics[width=85mm]{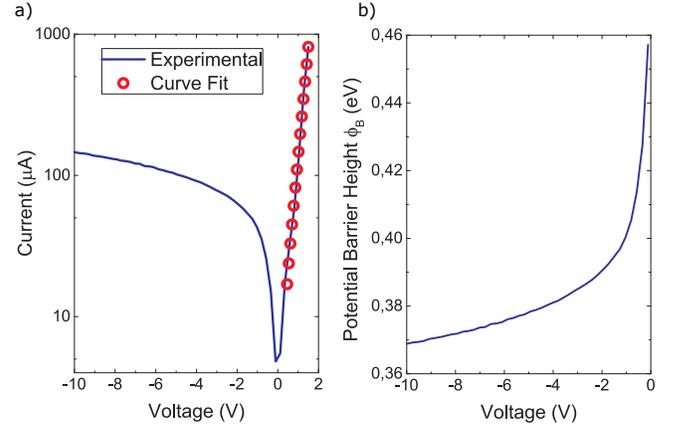}}
\caption{a) I-V characteristic of SLG/Si Schottky PD (semi-log scale). Experimental data and fit are shown. (b) Potential barrier height as a function of reverse bias.}
\label{fig:Fig5}
\end{figure}

To electrically characterize the PD, we measure the current-voltage (I-V) characteristics of the SLG/Si Schottky junction, Fig.\ref{fig:Fig5}a. The device shows rectifying I-V diode behavior, which follows the Schottky diode equation\cite{Sze2006, TongPRX2012}:
\begin{equation}
I= AA^*T^2e^{-\dfrac{\Phi _B}{k_B T}}\left(e^{\dfrac{qV_a}{\eta k_B T}}-1\right)
\label{eq1}
\end{equation}
where $\Phi_B= \Phi_{B0}+\Delta\Phi_B(V)$, $\Phi_{B0}$ is the Schottky barrier height (SBH) at zero voltage, $\Delta\Phi_B(V)$ is the SBH change due to applied voltage bias, \textit{A}$^*$ is the Richardson constant (32 A/cm$^2$K$^2$ for p-type Si\cite{Sze2006}), \textit{A} is the junction area, \textit{k$_B$T} is 26meV at room temperature, $\eta$ is the diode ideality factor, defined as the deviation of the measured I-V curve from the ideal exponential behavior\cite{Sze2006}, and \textit{V}$_a$ is the applied voltage. $\Delta\Phi_B(V)$ is typically dominant in reverse bias because of the higher potential drop on the Schottky junction, resulting in a more pronounced barrier-lowering Schottky effect\cite{Sze2006,TongPRX2012} and SLG Fermi level shift\cite{AnNL2013,TongPRX2012,BartoPR2016}. On the other hand, in forward bias the potential drop is limited by the built-in voltage (V$_{bi}<$1V)\cite{Sze2006}, so that $\Delta\Phi_B(V)$ can be neglected and $\Phi_B\sim \Phi_{B0}$. We estimate SBH in forward bias by fitting the experimental data with Eq.\ref{eq1} and using $\Phi_B$ and $\eta$ as fitting parameters. We get $\Phi_B\sim$0.46eV and $\eta\sim$11. These are in the range of previously reported values (0.41$<\Phi_B<$0.47 and 2$<\eta<$30) for SLG/Si Schottky diodes\cite{AmirIEEE2013,AnNL2013,WangNP2013,GoykNL2016,BartoPR2016, ChenNL2011}. By fitting the I-V curve in reverse bias we get the SBH dependence on applied reverse voltage and find $\Delta\Phi_B$ up to$\sim$80meV at -10V, Fig.\ref{fig:Fig5}b.
\begin{figure}
\centerline{\includegraphics[width=85mm]{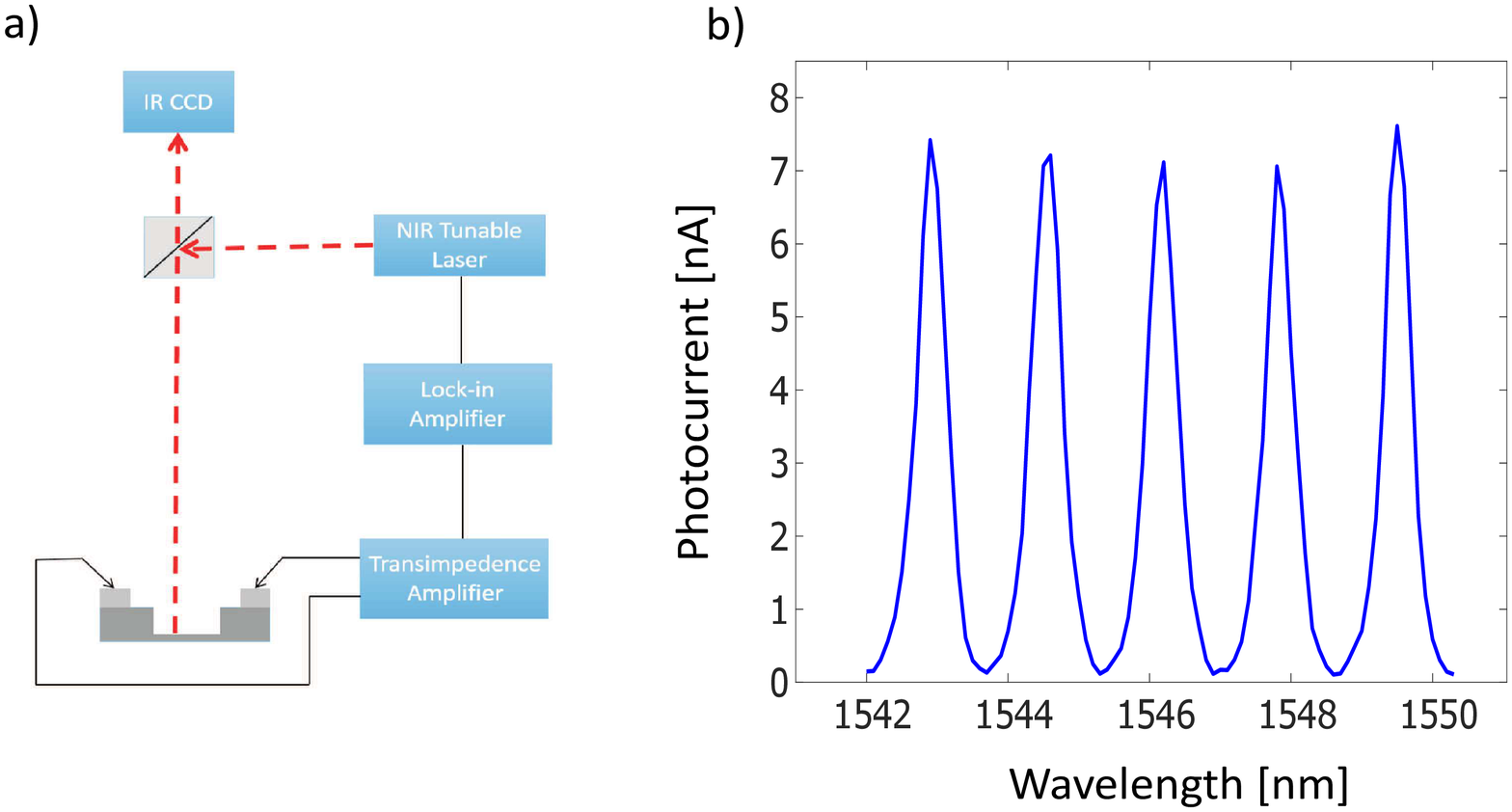}}
\caption{a) Opto-electronic measurements setup. (b) Spectral response (photocurrent) of RCE SLG/Si Schottky PD without backside Au (V$_R$=1V)}
\label{fig:Fig6}
\end{figure}

The opto-electronic characterizations is carried out using the set-up of Fig.\ref{fig:Fig6}a. The signal from a tunable laser (ANDO AQ4321D) is collimated, chopped and split between the reference, used for continuous power monitoring, and the device under test. The PD illumination is inspected by an IR camera. The photocurrent is amplified with a transimpedance amplifier (CVI Melles Griot 13AMP005) and fed to a lock-in amplifier (Signal Recovery 7280 SDP) for measuring the photoresponse. The incident optical power P$_{inc}$ is measured separately with a InGaAs PD (Thorlabs DET410). Fig.\ref{fig:Fig6}b plots the spectral response under reverse voltage V$_R$=1V. The device demonstrates spectral selectivity, exhibiting wavelength dependent and periodic photocurrent peaks upon illumination. The spectral separation between the peaks is$\sim$1.7nm, matching the FSR=1.65nm of the  F-P cavity. As expected, at resonance we get photocurrent peaks due to increased absorption at the Schottky interface, Fig.\ref{fig:Fig2}.

\begin{figure}
\centerline{\includegraphics[width=80mm]{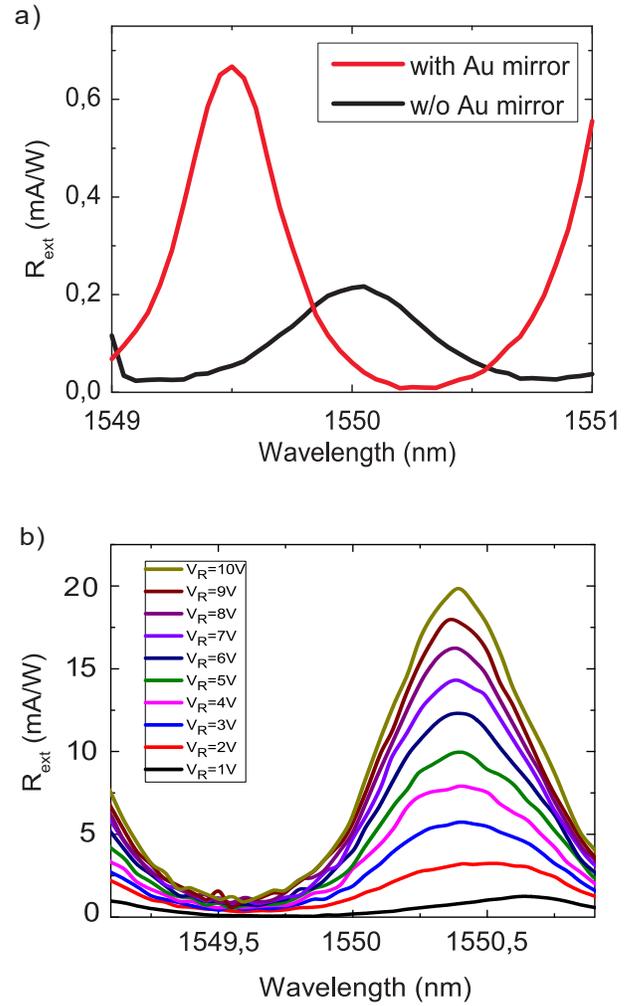}}
\caption{(a) R$_{ext}$ of a SLG-based RCE PD with (red curve) and without (black curve) Au mirror at -1V. (b) R$_{ext}$ as a function of increased reverse voltage}
\label{fig:Fig7}
\end{figure}

To confirm the cavity effect on PD responsivity, we measure $R_{ext}$ with and without Au mirror, Fig.\ref{fig:Fig7}a. The slight ($\sim$0.5nm) variation in resonant wavelength (Fig.\ref{fig:Fig7}a,b) is attributed to fabrication tolerances between different devices. We observe a 3-fold $R_{ext}$ enhancement with the Au mirror compared to the bare Si/air interface, in good agreement with the simulations in Fig.\ref{fig:Fig2}. To further enhance $R_{ext}$, we exploit the Schottky barrier lowering effect and apply a larger (up to 10V) reverse bias to the PDs with integrated Au mirrors. Fig.\ref{fig:Fig7}b plots $R_{ext}$ for different V$_R$. We get $R_{ext}\sim$20mA/W at V$_R$=10V, which corresponds to $R_{int}\sim$0.25A/W, considering the 8$\%$ absorption in the SLG electrode. To the best of our knowledge, this is the highest value reported so far for vertically-illuminated Schottky Si PDs at 1550nm. $R_{ext}$ may be further increased by integrating an antireflection coating to further enhance the SLG absorption.

In summary, we demonstrated a spectrally-selective, free-space illuminated SLG/Si Schottky PD at 1550nm. The photodetection mechanism is based on internal photoemission at the SLG/Si interface. The photodetection is enhanced by integration in a F-P cavity and increasing the SLG absorption due to multiple reflections at the cavity resonance. As a result, we showed wavelength-dependent photoresponse with external (internal) responsivity$\sim$20mA/W (0.25A/W). The resonance wavelength may be further tuned by varying the Si cavity thickness while the spectral-selectivity can be increased by taking advantage of more complex high-finesse microcavities, which would provide both integrated spectral filtering and enhanced SLG absorption. Our devices pave the way for developing high-responsivity graphene-Si free-space illuminated PDs for for NIR.

We acknowledge funding from EU Graphene Flagship, ERC Grant Hetero2D, EPSRC Grants EP/K01711X/1, EP/K017144/1, EP/N010345/1, and EP/L016087/1.

\end{document}